\documentclass[12pt]{JHEP3}
\usepackage{amsmath,amssymb}
\usepackage[latin1]{inputenc}

\def\G0{G_0}
\def\G1{G_{-1}}
\def\G3{G_{-\frac{1}{3}}}
\def\G23{G_{-\frac{2}{3}}}

\def\IT2{\int d^2\theta}
\def\bIT2{\int d^2\bar\theta}
\def\IT{\int d^2\theta d^2\bar\theta}

\newcommand{\spin}[6]
{ \Lambda^{\stackrel{#1}{#2}\stackrel{#3}{#4}
    \raisebox{0.7ex}{$\scriptscriptstyle #5 #6$} } }
\newcommand{\barspin}[6]
{ \Lambda^{\raisebox{0.7ex}{$\scriptscriptstyle #1 #2$}
    \stackrel{#3}{#4}\stackrel{#5}{#6}  } }

\title{$GSO(-)$ Vertex Operators and Open Superstring Field Theory in
  Hybrid Variables} \author{Luciano Barosi
  \thanks{lbarosi@fma.if.usp.br}\\
  \emph{Departamento de Física Matemática - IFUSP} \\\emph{C.P.66
    318, 05315-970, São Paulo, SP Brasil }}
  \author{Carlos Tello \thanks{telloc@wpunj.edu}\\\emph{Department
    of Mathematics, William Paterson University}\\
  \emph{Wayne, NJ 07470 USA}}

\preprint{IFUSP 1564/2003}  

\abstract{
  We consider the $GSO(-)$ sector of the open superstring using the
  formalism with four-dimensional hybrid variables. This sector is defined
  by the world sheet hybrid variables
  $(\theta^{\alpha},\bar{\theta}^{\dot\alpha})$ with antiperiodic
  boundary conditions. The corresponding spectrum of states and conditions
  for physical vertices are described. In particular we construct
  explicitly the lower level $GSO(-)$ vertex operators corresponding to
  the tachyon and the massless fermions.  Using these new vertices, the
  tachyon and massless sector contribution to the superstring field theory
  action of Berkovits is evaluated. In this way we have included the Ramond sector
  and we end by discussing some features of the action.
}

\begin{document}
\section{Introduction}
It has been considered that the more complete framework to study
the dynamics of the unstable non-BPS D-branes must be based on
open superstring field theory \cite{Sen1}. This theory must
include both Gliozzi-Scherk-Olive $(GSO)$ sectors of the
superstring that means the standard positive $GSO(+)$ and the
negative $GSO(-)$ projected sector. The process of annihilation of
these unstable non-BPS D-branes was studied using superstring
field theory action involving only the Neveu-Schwarz $(NS)$
sector, without the Ramond $(R)$ sector. However, it is still not
clear what is the role of the spacetime supersymmetry during this
process. For this reason, it is very interesting to explore some
aspects of this process from the point of view of spacetime
supersymmetry, as in \cite{Yoneya2}. Since the spacetime
supersymmetry always involves the contribution of both the $NS$
and $R$ sectors of the superstring we need to extend the framework
to keep track of the Ramond sector.

There is a formulation of the superstring in terms of hybrid
variables which is suitable to explore the issues of space-time
supersymmetry. Specifically, we will use the four-dimensional
version of this formalism to study both $GSO$ sectors and both $R$
and $NS$ sectors of the superstring that means we consider: $NS+$,
$NS-$, $R+$ and $R-$ sectors.

The standard $GSO(+)$ sector has spacetime supersymmetry and it is
identified with the periodic sector of the $\theta,\bar{\theta}$
world sheet variables. In this sector all vertex operators are N=1
D=4 superfields due to the fact that the spacetime supersymmetry
current has zero modes. A detailed discussion of this sector was
given in references \cite{Berkovits4} of Berkovits.

On the other hand, the $GSO(-)$ sector is not spacetime
supersymmetric and it is identified with the antiperiodic sector
of the $\theta, \bar{\theta}$ world sheet variables
\cite{Yoneya2}. In this sector the vertex operators are ordinary
fields and the spacetime supersymmetry is broken since the
supercurrents have half integer modes.

Using these ideas in order to understand both GSO sectors, we
proceed as follows:

In section 2 we review the hybrid approach for superstring and
explain its relation with the RNS approach. We use the mapping
between the two sets of variables to show the vertex operators for
tachyon and massless sector in both formalisms. The conditions for
physical vertices are also discussed.

In section 3, as a consistency check, we construct the physical
$GSO(-)$ vertices that correspond to the tachyon and the massless
fermions using the hybrid variables. All these vertices contain
spin fields to change to antiperiodic boundary conditions for the
${\theta}^{\alpha},\bar{\theta}^{\dot\alpha}$ variables.

In section 4 we review the open superstring field theory action
which describes non-BPS D-branes using the hybrid variables. This
action requires three independent superstring fields represented
by $\Phi$, $\Psi$ and $\bar{\Psi}$ which are appropriately
tensorized, so as to take into account  the two GSO projections.

In section 5, we explicitly compute the classical superstring
field theory action including the contribution of the tachyon, the
massless fermions, and massless gauge bosons.

The full action, written using explicit 10-dimensional Lorentz invariant
fields is \footnote{Actually, the term with 4 gauge fields in
  $F^{mn}F_{mn}$ has never been calculated and most likely is not going to
  give the correct coefficient without going to an effective action.
  Anyway, we don't use this result anywhere in this work.}:
\begin{eqnarray}
  \nonumber S&=& \mathrm{Tr} \int d^{10} x
\left[ \frac{1}{4} F^{\mu\nu}F_{\mu\nu}+ T\Box T +\frac{1}{2}T^2
  +[T,A^{\mu}][T,A_{\mu}]-T^4
\right. \nonumber \\
&& + \chi^{a\;+} \gamma^{\mu}_{ab} \left(\partial_\mu
\chi^{b\;+}+[A_{\mu},\chi^{b\;+}] \right) + \chi^{a\;-}
\gamma^{\mu}_{ab} \left(\partial_\mu
\chi^{b\;-}+[A_{\mu},\chi^{b\;-}] \right)
\nonumber\\
&&\left. + 3^\frac{3}{4} T (\chi^{a\;+} \chi^-_{a})+\cdots \right]
\end{eqnarray}
where $F_{mn}$ is the field-strength for the gauge-field $A_m$,
$T$ is the tachyon field, $\chi^{+}_{a}$, $\chi^-_{a}$ are the
massless fermions coming from the $GSO(+)$ and $GSO(-)$ sector
that are 10-dimensional chiral and antichiral respectively.

At the end we comment on some of the features of this action.

\section{Hybrid Superstring with both $GSO(+)$
  and $GSO(-)$ sectors}

In this section we review the relation between RNS and hybrid
variables. Using the map between them we show the vertex operators
in both formalisms and we discuss the conditions for physical
vertex operators.

\subsection{RNS Variables}

First we represent the RNS Matter and ghost variables in the large
Hilbert space:
\begin{equation}
[\; \underbrace{x^{m}, \; \psi^{m},\;b,\; c,\; \xi,\;\eta,\;
\phi}_{\mbox{{\tiny $D=4$ space-time matter + ghosts}}} ,\;\;
\underbrace{ x^{\pm j},\; \psi^{\pm j}}_{CY} \;], \label{2}
\nonumber
\end{equation}
the first set of describes the RNS $D=4$ spacetime matter plus the
ghost worldsheet fields and the second set describes the other six
dimensional internal worldsheet fields. These last internal fields
may be compactified in a Calabi-Yau or not, but in the present
case we just complexify them as explained later on in this same
paragraph. Specifically we have: $ x^m:\; (m=0,1,2,3),$ are the
$D=4$ spacetime vector RNS world sheet bosonic coordinates; $
x^{\pm j} = \frac{1}{\sqrt{2}}(x^{2j+2}\pm i x^{2j+3}),\;
(j=1,2,3)$. are the internal d=6 RNS world sheet bosonic
coordinates; $\psi^m,\;\; (m=0,1,2,3),$ are the $D=4$ spacetime
vector RNS world sheet fermionic coordinates; $\psi^{\pm j} =
\frac{1}{\sqrt{2}}(\psi^{2j+2}\pm i\psi^{2j+3}),\;\;(j=1,2,3)$ are
the internal d=6 RNS world sheet fermionic coordinates. See
appendix A about the properties of these RNS variables.

In order to construct space time fermion vertex operator, we need
to define the $D=4$ spacetime RNS spin fields $\Sigma^{\alpha}$
and $\bar{\Sigma}^{\dot\alpha}$ by
\begin{eqnarray}
\Sigma^{\alpha} &=& \Sigma^{\pm\pm} \;\;(even
\;+),            \nonumber\\
\bar{\Sigma}^{\dot\alpha} &=& \Sigma^{\pm\pm};\;\;(odd
\;+).\nonumber
\end{eqnarray}
and the U(1) internal chiral boson $H$ by
\begin{eqnarray}
J &=& \partial H =i\left(\psi^4\psi^5 + \psi^6\psi^7  +
   \psi^8\psi^9\right)=:\psi^{+j}\psi^{-j}:,
\nonumber\\
   H(z)H(w) & = & 3\ln(z-w).\nonumber
\end{eqnarray}
As is well known, the RNS approach of superstring has the problem
that the off-shell space-time supersymmetry is not manifest due to
the problem of picture \cite{fms}. This problem makes it difficult
to set up a second quantized approach of superstrings which is
gauge invariant and free of contact term divergences at the tree
level at the same time. However, as we will discuss in the next
subsection, there is a more convenient formulation of the
superstring. This alternative formulation is the so called hybrid
approach, and preserves off-shell supersymmetry.

\subsection{Hybrid Variables}
In the hybrid approach of the superstring the fundamental
variables are the following world sheet fields:
\begin{equation}
\left[\;\; X^m,\;\theta^{\alpha},\;\bar\theta^{\dot\alpha},\;
p_{\alpha},\; \bar{p}_{\dot\alpha}, \;\rho,\;\Gamma^{+j},
\Gamma^{-j},\; X^{+j},\; X^{-j}\;\; \right] \label{3} \nonumber
\end{equation}
here $X^m$ is the $D=4$ spacetime vector world
sheet boson, $m=0,1,2,3$;\\
$\theta^{\alpha}$ is the $D=4$ spacetime chiral spinor worldsheet
fermion, $\alpha=1,2$; \\
$\bar{\theta}^{\dot\alpha}$ is the space-time anti-chiral spinor
world
sheet fermion, $\dot\alpha=1,2$; \\
$\rho$ is the world sheet chiral boson;\\
$\Gamma^{\pm j}$ are internal world sheet
fermions, $j=1,2,3$; \\
$X^{\pm j}$ are internal world sheet bosons,
$j=1,2,3$.\\

The hybrid variables are defined in such a way that we have two
independent set of worldsheet fields. The first set satisfy a
$N=2,\; c=-3$ superconformal algebra and describes the $N=1$,
$D=4$ super spacetime and the second set satisfy a $N=2,\; c=9$
superconformal algebra and describes the internal coordinates.
Even though any CY compactification is allowed for this last set
of fields, we are going to consider only the complexification of
them in flat space.

In the Appendix B, all the properties of the hybrid variables are
given. And we should remember that the mapping between RNS and
hybrid variables is a unitary transformation as in reference
\cite{Berkovits4}

The non-vanishing OPEs for the hybrid variables are:
\begin{eqnarray}
X^m(z)X^n(w)   & = & -\eta^{mn} \ln(z-w),
\nonumber\\
p_{\alpha}(z)\theta^{\beta}(w) & = &
\frac{\delta^{\beta}_{\alpha}}{z-w}, \nonumber
\\
\rho(z)\rho(w) & = & -\ln(z-w), \nonumber\\
\Gamma^{+i}(z) \Gamma^{-j}(w) & = &
\frac{\delta^{ij}}{(z-w)}, \\
X^{+i}(z)X^{-j}(w) &=& -\delta^{ij}\ln(z-w).\label{opes}\nonumber
\end{eqnarray}

The hybrid superstring is described by the following quadratic
world sheet action
\begin{eqnarray}
S_{hyb} &=&  \int d^2 z \Big[
 \frac{1}{2}\partial X^m \bar\partial X_m
 + p_{\alpha} \bar\partial \theta ^{\alpha}
+ \hat p_{\dot\alpha} \bar \partial
 \hat\theta^{\dot\alpha}+ \bar{p}_{\alpha}
\partial \bar{\theta}^{\alpha} + \bar{\hat{
p}}_{\dot\alpha}
\partial\bar{\hat{\theta}}^{\dot\alpha}\Big]+
S_{\rho}
 + S_c, \nonumber
\end{eqnarray}
where $S_{\rho}$, $S_c$ are the actions for the chiral $\rho$
boson and the compact variables respectively.

This world sheet action has a $N=2$ superconformal algebra formed
with the generators $L$, $G^+$, $G^-$, $J_{gh}$. It turns out that
is convenient to extend this superconformal algebra to form an
$N=4$ superconformal algebra formed by the following generators:
$J^{--}$ with conformal weight $0$; $G^+,\;\tilde{G}^+,\; J_{gh}$
with conformal weight $1$ and $L,\;G^-,\;\tilde{G}^-,\;J^{++}$
with conformal weight 2 (here the upper indices $+$ and $-$
indicate the $U(1)$ ghost charge generated by $J_{gh}$):
\begin{eqnarray}
\label{n=4} L &=& -\frac{1}{2} \partial X^m\partial X_m
-p_{\alpha}\theta^{\alpha} -
{\bar{p}}_{\dot\alpha}{\bar\theta}^{\dot\alpha}
-\partial\rho\partial\rho -\frac{1}{2}\partial^2\rho
+L_c,\nonumber
\\
G^+&=& \underbrace{e^{\rho}(d)^2}_{G^+_0} +
\underbrace{\Gamma^{-j}\partial X^{+j}}_{G^+_{-\frac{1}{3}}},
\nonumber \\
\tilde{G}^+&=& \underbrace{\epsilon^{jkl} \Gamma^{-j}
\Gamma^{-k}\Gamma^{-l}e^{-2\rho}(\bar d)^2}_{G^+_{-1}} +
\underbrace{e^{-\rho}\epsilon^{jkl} \Gamma^{-j}\Gamma^{-k}\partial
X^{-l}}_{G^+_{-\frac{2}{3}}},
\nonumber\\
G^-&=& \underbrace{e^{-\rho}(\bar d)^2}_{G^-_{0}}
+\underbrace{\Gamma^{+j}\partial X^{-j}}_{G^-_{\frac{1}{3}}},\\
\tilde{G}^-&=& \underbrace{\epsilon^{jkl} \Gamma^{+j}
\Gamma^{+k}\Gamma^{+l}e^{+2\rho}( d)^2}_{G^-_1} +
\underbrace{e^{\rho}\epsilon^{jkl} \Gamma^{+j}\Gamma^{+k}\partial
X^{+l}}_{G^-_{\frac{2}{3}}},     \nonumber\\
J_{gh} &=& \partial \rho +
\Gamma^{-j}\Gamma^{+j},       \nonumber \\
J^{++} &=& e^{-\rho-\Gamma^{-j}\Gamma^{+j}},
   \nonumber\\
J^{--} &=& e^{\rho+\Gamma^{-j}\Gamma^{+j}},
      \nonumber
\end{eqnarray}
where $ L_c$ is the stress tensor for the internal variables
$\Gamma^{\pm
  j},\;X^{\pm j}$ and
\begin{eqnarray}
d_{\alpha}(z)&=&p_{\alpha}(z)+
\frac{i}{2}\bar{\theta}^{\dot\alpha}
\sigma^m_{\alpha\dot\alpha}\partial X_m(z) -\frac{1}{4}(\bar
\theta)^2\partial\theta_{\alpha}(z)
+\frac{1}{8}\theta_{\alpha}\partial(\bar\theta)^2(z) \label{d}
\end{eqnarray}
is the current associated to the $N=1$, $D=4$ supersymmetric
covariant derivative $D_{\alpha}$.

In the equations (\ref{n=4}) we have introduced a notation
suitable for writing the superstring field theory action we are
going to deal with in section 4, labelling each term in the
operators $G$ and $\tilde G$ by a subscript denoting its C-charge,
defined by
\begin{equation}
  \label{c-charge}
  C=\frac{1}{3}\int \Gamma^{+j}\Gamma^{-j}.
\end{equation}
This charge is related to the ghost number and $\rho$-charge, as
it is obvious from the expression of $J_{gh}$ in eq.(\ref{n=4}). A
detailed account of the construction of the superstring action and
the role of the C-charge in this construction may be found in
reference \cite{Berkovits7}.

As we will discuss in detail on the next section about vertex
operators, there are two sectors in the hybrid approach of
open superstring:\\
The sector with periodic boundary conditions in the fermionic
variables ($\theta,\;p,\;\bar\theta,\;\bar{p}$) is identified with
the $GSO(+)$ sector of the open superstring.  This sector has
manifest $N=1,\;D=4$ spacetime supersymmetry and was studied in a
series of papers by Berkovits \cite{Berkovits4}. It is remarkable
that this approach contains the four dimensional contribution of
both the $NS(+)$ and $R(+)$ sub-sectors through vertex operators
that are explicitly contained in $N=1$, $D=4$ spacetime
superfields. This fact is an important advantage compared with the
RNS approach where this is not possible.

The other sector is defined by anti-periodic boundary conditions
in all the fermionic variables ($\theta,\;
p,\;\bar\theta,\;\bar{p}$) is identified here with the $GSO(-)$
sector of the open superstring. Since there is no fermionic zero
modes in this sector the $N=1,\;D=4$ spacetime supersymmetry is
completely broken. Furthermore, as we will show in detail, this
sector contain the four-dimensional contribution of both the
$NS(-)$ and $R(-)$ sub-sectors.

\subsubsection{Hybrid Spin Field Operator} In order to construct
the anti-periodic sector of the fermionic hybrid variables
$\theta,\;\bar{\theta},\;p,\bar{p}$ we need a spin field operator.
Firstly, we have to bosonize these variables as follow:
\begin{equation}
\begin{array}{llll}
\theta^1 = e^{\sigma_1},& \theta^2 = e^{\sigma_2}, &
\bar{\theta}^1= e^{\bar{\sigma}_1},& \bar{\theta}^2 =
e^{\bar{\sigma}_2},          \nonumber \\
p_1 = e^{-\sigma_1},& p_2 = e^{-\sigma_2},& \bar{p}_1 =
e^{-\bar{\sigma}_1},& \bar{p}_2 = e^{-\bar{\sigma}_2},
\end{array}
\end{equation}
where the chiral bosons
$\sigma_1,\;\sigma_2,\;\bar{\sigma}_2,\;\bar{\sigma}_1$ satisfy
the OPEs:
\begin{equation}
\begin{array}{lcl}
\sigma_1 (z) \sigma_1 (w) = \ln(z-w),&\;& \sigma_2 (z) \sigma_2
(w) = \ln(z-w),
\nonumber\\
\bar{\sigma}_1 (z) \bar{\sigma}_1 (w) = \ln(z-w),
 &\;& \bar{\sigma}_2 (z) \bar{\sigma}_2 (w) =
\ln(z-w).
\end{array}
\end{equation}
In general, all vertex operators of the $GSO(-)$ sector must
contain the following general hybrid spin field
\begin{eqnarray}
\Lambda^{A_1A_2B_1B_2}&=& e^{A_1\sigma_1 + A_2\sigma_2 +
B_1\bar{\sigma}_1 + B_2\bar{\sigma}_2}. \nonumber
\end{eqnarray}
where $ A_1, A_2, B_1, B_2 $ are all half integer numbers and the
conformal weight is
\begin{equation}
W[\Lambda^{A_1A_2B_1B_2}]
=\frac{1}{2}[A_1(A_1-1)+A_2(A_2-1)+B_1(B_1-1)+B_2(B_2-1)].
\end{equation}
We will use as shorthand index notation

$(+=\frac{1}{2}),\;(-=-\frac{1}{2}),\;(\oplus=\frac{3}{2}),
\;(\ominus=-\frac{3}{2}),\;(\boxplus=\frac{5}{2}),\;(\boxminus=-\frac{5}{2})$.
This construction is in some sense analogous to the construction
of the Ramond vacuum in the RNS superstring theory \cite{fms}.
Here we define $|\Lambda\rangle$ as the vacuum state for the
anti-periodic variables $p\theta$, $\bar{p}\bar{\theta}$
\begin{equation}
|\Lambda\rangle = \Lambda^{++++}|0\rangle =
:e^{(\frac{1}{2}\sigma_1+\frac{1}{2}\sigma_2+\frac{1}{2}\bar{\sigma}_1
  +\frac{1}{2}\bar{\sigma}_2)}(0):|0\rangle.
\end{equation}
This $|\Lambda\rangle$ vacuum is non degenerate and it is
annihilated by all fermionic oscillators with positive modes.
$\theta^{\alpha}_r|\Lambda\rangle =p^{\alpha}_r|\Lambda\rangle
=0,\;\;\; r=\frac{1}{2},\frac{3}{2},\frac{5}{2},\cdots etc.$

\subsection{Hybrid Variables vs RNS Variables
  Mapping} The hybrid variables are related with the standard RNS
variables by the following field redefinition \cite{Berkovits4}:
\begin{eqnarray}
X^m &=& e^{R+\frac{1}{2}U} (x^m)e^{-R-\frac{1}{2}U},
\nonumber
\\
\theta^{\alpha} & = & e^{\frac{\phi}{2}}\Sigma^{\alpha}
e^{-\frac{H}{2}},  \nonumber  \\
\bar{\theta}^{\dot{\alpha}} & = & c\xi
e^{-\frac{3}{2}\phi}\Sigma^{\dot\alpha} e^{\frac{H}{2}},
     \nonumber\\
p_{\alpha} &=& e^{R+\frac{U}{2}}
(e^{-\frac{\phi}{2}}\Sigma_{\alpha} e^{\frac{H}{2}})
e^{-R-\frac{1}{2}U},
       \nonumber\\
\bar{p}_{\dot\alpha} &=& e^{R+\frac{1}{2}U} (b\eta
e^{\frac{3\phi}{2}}\Sigma_{\dot\alpha} e^{-\frac{H}{2}})
e^{-R-\frac{1}{2}U},
      \nonumber\\
\partial \rho & = & 3\partial\phi - cb -
2\xi\eta -\partial H,
      \nonumber\\
X^{+j} &=& e^{R+\frac{1}{2}U} ( x^{+j})e^{-R-\frac{1}{2}U},
       \nonumber\\
X^{-j} &=& e^{R+\frac{1}{2}U} ( x^{-j})e^{-R-\frac{1}{2}U},
      \nonumber\\
\Gamma^{+j} &=& e^{R+\frac{1}{2}U} ( \xi
e^{-\phi}\psi^{+j})e^{-R-\frac{1}{2}U},
      \nonumber\\
\Gamma^{-j} &=& e^{R+\frac{1}{2}U} ( \eta
e^{\phi}\psi^{-j})e^{-R-\frac{1}{2}U},
\end{eqnarray}
where
\begin{eqnarray}
R &=& \int dz \;c \xi e^{-\phi}i\psi^{-j}\partial
x^{+j},                  \nonumber \\
U&=& \int dz [\; c\xi\; e^{-\phi}i\psi^m\partial x_m
+\frac{1}{2}(\partial\phi+\partial H)c\partial c \xi\partial\xi
e^{-2\phi}]).\nonumber
\end{eqnarray}

This mapping ensures that Eq.(2) are the complete set of
non-vanishing OPEs of the hybrid variables.

\subsection{Conditions for Physical Vertex  Operator}

String fields are vertex operators acting on the Hilbert space of
the string. In this section we are going to show the properties of
these operators, stressing what conditions they have to satisfy to
be physical.

As usual in string theory, we have several different ways of
stating these physical conditions. Even though they are seemingly
different, they are all equivalent. We are going to present three
different possibilities. In a later section, the second and third
set of conditions will be used and shown to yield the same result
coming from the RNS conditions. We also argue on the equivalence
of the different approaches.

\newpage
 \centerline{\bf Vertex Operators in RNS formalism}
 A physical RNS vertex $V$ has the following
features:
\begin{itemize}
\item V is in the cohomology of the BRST operator $Q$, that is:
\begin{equation*}
 \{Q,V\} = 0, \qquad V \ne Q \Lambda.
\end{equation*}
Note that $V$ and $\Lambda$ are fields in the small Hilbert space
(which does not contain $\xi$ zero modes). \item Each vertex V
comes from a definite GSO sector, either GSO$(+)$ or GSO$(-)$. The
former being $NS+$ (fermionic) and $R+$ (bosonic), the latter
being $NS-$ (bosonic) and $R-$ (fermionic). \item The vertex V
must have ghost number $+1$. \item V must have a definite picture.
Usually the picture is chosen to be $-1$ for the $NS+$ sector and
$-\frac{1}{2}$ for the $R+$ sector.
\end{itemize}

In the hybrid formalism we have two setups to consider in order to
understand physical states, one coming from the $N=2$ superalgebra
and the other from the $N=4$ superalgebra.

The conditions for the $N=4$ superalgebra are easier to understand
since the conditions may be translated directly from the RNS
formalism.

\vspace{0.5cm} \centerline{\bf $N=4$ Physical Vertex Operators}
\vspace{0.3cm}

In order to understand the equivalence between these conditions
and the RNS conditions, remember that the operators $G^+$ and
$\tilde G^+$ are the BRST-current and the $\eta$ ghost
respectively. So, we may directly translate the conditions for the
vertices to be in the small Hilbert space and the BRST conditions
as the set of equations $\tilde G^+ V=0$, $G^+ V=0$. Solving the
first equation (due to the triviality of the cohomology of $\eta$)
we have $V=\tilde G^+ \Phi$, where $\Phi=\xi V$. $G^+$ constraint
yields the correct conditions. The only subtlety is the gauge
invariance, which can be understood in the same way. The
conditions on statistics are reversed due to the presence of the
$\xi$ fields in the vertices. Finally, picture is related to the
choice of $\rho$-charge by the relation $P=\frac{1}{2\pi}\int dz
(i\partial\rho -\frac{1}{2}p_ \alpha \theta^\alpha +
\frac{1}{2}\bar p_{\dot\alpha} \bar\theta^{\dot\alpha})$. We refer
the reader to the reference \cite{Berkovits7} for a discussion of
this choice and the need for three different values for the
picture number P and rho charge $\rho$. So, the condition for
$N=4$ physical vertices are:
\begin{itemize}
\item We should consider vertices in the large Hilbert space,
$\Phi=\xi V$. For these kind of vertices the BRST condition of the
RNS translates to:
\begin{equation*}
G^+ \tilde G^+ \Phi =0, \qquad \Phi\ne G^+ \Lambda + \tilde G^+
\tilde \Lambda.
\end{equation*}
These are also the linearized equations of motion coming from the
superstring field theory action. \item Again the vertices come
from one of the two GSO sectors, but their statistics are now
reversed with respect to the RNS. \item Ghost charge is the $U(1)$
charge of the $N=4$ formalism, and we should demand  all the
vertices to have zero ghost charge. \item Choosing a picture in a
RNS formalism is the responsible for the lack of manifest
supersymmetry. In the hybrid variables we may choose the values of
the $\rho$ charge instead. We choose $\rho$ charge zero for the
four-dimensional fields and $\rho$-charge $\pm1$ for the internal
6-dimensional fields. Another equivalent possibility is to choose
the value of the C-charge (defined in eq.(\ref{c-charge}) to have
one of three values in the set $\{-1/3,0,1/3\}$.
\end{itemize}

One should note that both set of conditions above allow us to
gauge transform the vertex operators. The next set of conditions
will be cast in a gauge-fixed form.

\vspace{0.5cm} \centerline{\bf $N=2$ Formalism} \vspace{0.5cm}

In this formalism we need to demand the fields to be $N=2$
superconformal primary fields with zero conformal weight and zero
$U(1)$-charge, which mean that they have to satisfy the following
conditions:

\begin{enumerate}
\item  $V$ must have zero conformal weight; the residue of double
pole
  (and higher) in the OPE of $L$ and $V$ must be zero.
\item  $V$ must satisfy the $G^+$ constraint; the residue of
double
  pole (and higher) in the OPE of $G^+$ and $V$ must be zero.
\item  $V$ must satisfy the $G^-$ constraint; the residue of
double
  pole and (higher)in the OPE of $G^-$ and $V$ must be zero.
\item  $V$ must have zero ghost charge;
  the residue of simple poles (and higher) in the OPE of $J_{gh}$ and $V$ must be zero.
\end{enumerate}

The issues about GSO projection and the choice of $\rho$ charge
are the same as in the $N=4$ formalism.

In order to understand the above conditions, we must remark that
the physical vertices, as usual, should be the operators in the
cohomology of the $N=2$ BRST charge which is constructed using the
superconformal generators $(L,\;G^+,\;G^-,\;J)$ plus four systems
of ghosts, for reparametrization, two dimensional $N=2$ SUSY and
$U(1)$ charge. However, in analogy to what happens in the RNS
formalism, one may show that $N=2$ superconformal primary fields
with zero weight and zero charge are in the cohomology of
$Q_{N=2}$. This is done by considering a particular explicit
dependence of the ghosts for the operators, which, just like in
the RNS formalism, amounts to choosing a kind of Feynmann-Siegel
gauge and imposing the vertices to be annihilated by the ghosts
$b$ and $\tilde b$, for reparametrization and $U(1)$ charge
\cite{Lechtenfeld}.

The next step to show the equivalence between the $N=2$ and the
RNS description is to show that the two cohomologies agree. This
was done in reference \cite{Petersen} by showing, via a similarity
transform, that the $N=2$ BRST charge may be written as a sum of
the $N=1$ BRST charge, a topological BRST charge and the $U(1)$
charge, all nilpotent and anticommuting with each other. The
conditions that the two last BRST charges impose are just
constraints in the ghost-dependence done in such a way that we can
construct the vertices without the need of the $N=2$ ghosts, using
only the $N=1$ matter and the $N=1$ ghosts.

\subsection{Vertex Operators}

In this subsection we present the vertex operators for the two
first mass levels, both in the RNS and in the hybrid variables.
The RNS vertices are easily found in the literature
\cite{Polchinski}, and we may translate them to the hybrid
formalism using the mapping discussed in the beginning of this
section.

The first thing we are going to do is to show the results in a
table, in order to assure the important results to be all
collected in a single place. After this, we explain the notation
and make a few comments.

\centerline{Table 1: Vertex Operators corresponding to the tachyon
and the massless fields}
\[
\begin{array}{|c|l|l|c|l|c|}
\hline
Vertex & \ \ \ \ \ \ \ \ RNS & \ \ \ \ Hybrid & W & S &  GSO\\
\hline
       & c\partial c \xi\partial\xi
         e^{-\frac{5}{2}\phi}\Sigma^{\alpha}e^{H/2}\chi^+_{\alpha}
       & \theta^{\alpha}(\bar{\theta})^2 \chi^+_{\alpha}
       & 0 & Ff & +\\
\cline{2-6}
       & c\xi
         e^{-\frac{1}{2}\phi}\Sigma^{\dot{\alpha}}e^{-H/2}
         \bar{\chi}^+_{\dot\alpha}
       & \bar{\theta}^{\dot\alpha} (\theta)^2\bar{\chi}^+_{\dot\alpha}
       & 0 & Ff & +\\
\cline{2-6} \Phi_0 & c\xi e^{-\phi}\psi^m A_m
       & \theta^{\alpha}\sigma^m_{\alpha\dot\alpha}
         \bar{\theta}^{\dot\alpha} A_m
       & 0 & Bb & +\\
\cline{2-6}
       & c\xi e^{-\phi}t
       & \Lambda t
       & -\frac{1}{2} & Fb & -\\
\cline{2-6}
       & z^{\frac{1}{2}}c\xi
         e^{-\frac{1}{2}\phi}\Sigma^{\alpha}e^{-H/2}\chi^-_{\alpha}
       & \theta^{\alpha} \Lambda\chi^-_{\alpha}
       & 0 & Bf & -\\
\cline{2-6}
       & z^{\frac{1}{2}}c\partial c \xi \partial\xi
         e^{-\frac{5}{2}\phi}\Sigma^{\dot\alpha}e^{H/2}
         \bar{\chi}^-_{\dot\alpha}
       & \bar{\theta}^{\dot\alpha} \Lambda \bar{\chi}^-_{\dot\alpha}
       &  0 & Bf & -\\
\hline
       & c\xi e^{-\frac{1}{2}\phi} \Sigma^{\alpha} \Xi^{+j}
         \lambda^+_{+j\alpha}
       & e^{\rho}\Gamma^{+j}\theta^{\alpha}(\bar{\theta})^2
         \lambda^+_{+j\alpha}
       & 0 & Ff & +\\
\cline{2-6} \Psi_{\frac{1}{3}}
       & c\xi e^{-\phi}\psi^{-j}A_{-j}
       & e^{\rho}\Gamma^{+j}(\bar{\theta})^2 A_{-j}
       & 0 & Bb & +\\
\cline{2-6}
       & z^{\frac{1}{2}}c\xi
         e^{-\frac{1}{2}\phi}\Sigma^{\dot{\alpha}}\Xi^{+j}
         \bar{\lambda}^-_{+j\dot\alpha}
       & e^{\rho} \Gamma^{+j} \bar{\theta}^{\dot\alpha} \Lambda
         \bar{\lambda}^-_{+j\dot\alpha}
       & 0 & Bf &-\\
\hline
       & c\partial c \xi\partial \xi
         e^{-\frac{5}{2}\phi} \Sigma^{\dot\alpha} \Xi^{-j}
         \bar{\lambda}^+_{-j\dot\alpha}
       & e^{-\rho} \Gamma^{-j} \bar{\theta}^{\dot\alpha} (\theta)^2
         \bar{\lambda}^+_{-j\dot\alpha}
       & 0 & Ff & +\\
\cline{2-6} \bar\Psi_{-\frac{1}{3}}
       & c\xi e^{-\phi}\psi^{+j}A_{+j}
       & e^{-\rho}\Gamma^{-j}(\theta)^2A_{-j}
       & 0 & Bb & +\\
\cline{2-6}
       & z^{\frac{1}{2}}c\partial c \xi\partial\xi
         e^{-\frac{5}{2}\phi}\Sigma^{\alpha}\Xi^{-j}\lambda^-_{-j\alpha}
       & e^{-\rho}\Gamma^{-j} \theta^{\alpha} \Lambda\lambda^-_{-j\alpha}
       & 0 & Bf & -\\
\hline
\end{array}
\]
\\

The first column denotes the different string fields, labeled by
their C-charges as defined in eq.(\ref{c-charge}). The second and
third columns show the expression for the different fields in RNS
and in Hybrid variables, respectively. In order to understand how
the fields are split one should have in mind the definitions for
the world-sheet fields [given in section (2.1) and (2.2)] and note
that the spectrum of the superstring (with both $GSO$ sectors) in
the two lowest mass-levels contains a tachyon, a gauge field and
two massless fermions of opposite chiralities. Since we want to
consider fields whose 4-dimensional transformations are explicit,
this means that we are considering a subgroup of the full
10-dimensional Lorentz group:

\begin{eqnarray}
\label{lorbreak}
  \nonumber \mathrm{SO(9,1)}&\rightarrow&
\mathrm{SO(3,1)}\times \mathrm{SO(6)},\\
  \mathrm{SO(9,1)}&\rightarrow&
\mathrm{SO(3,1)}\times
  \mathrm{SU(3)}\times \mathrm{U(1)}.
\end{eqnarray}

The last line comes from the fact that we need to consider
spinors. The representations of the $D=10$ Lorentz algebra
decompose as:
\[
\begin{array}{ccccccc}
  10 &\rightarrow & 4     & \oplus &   3     &
\oplus & \bar 3 \\
   A_{\mu}  &            & A_m        &&    A_{-i}  &
      &  A_{+i}
\end{array}
\]
 \[
 \begin{array}{ccccccccc}
  16 &\rightarrow& (2,1_{+3})& \oplus &
(2,3_{-1}) & \oplus & (2', 1_{-3}) &\oplus&
(2',\bar 3_{+1})\\
    \chi^+_{a} &           & \chi^+_{\alpha} &       &
\lambda^+_{-i\alpha}&&
  \bar{\chi}^+_{\dot\alpha} &&
\bar{\lambda}^+_{+i\dot\alpha} \\
  &&&&&&&&\\
  16' &\rightarrow& (2',1_{+3})
&\oplus&(2',3_{-1})& \oplus& (2, 1_{-3})
&\oplus & (2,\bar 3_{+1})\\
  \chi^-_{a}&& \bar{\chi}^-_{\dot\alpha} &&
\bar{\lambda}^-_{-i\dot\alpha} &&
  \chi^-_{\alpha} && \lambda^-_{+i\alpha}.
\end{array}
\]

In the above equations, primed quantities indicate anti-weyl
representations of the appropriate Lorentz group and unprimed
represent the weyl representations. The number in parenthesis mean
the dimension of the representations and the subscript is the U(1)
charge. Below the representation we indicate the fields which
transform under them. With this splitting, one can understand the
origin of all the spacetime fields appearing in the table 1.

Also on the table 1,  the $W$ column shows the conformal weight of
the fields. The $S$ column represents both the  statistics of the
spacetime fields, $b$ stands for spacetime boson and $f$ for
spacetime fermionic; the capital letter denote the statistics in
the world-sheetfield that means: $B$ stands for world sheet
bosonic field and $F$ stands for world sheet fermionic field.
Finally, the last column represents the GSO sector originating the
fields.

In the next section we are going to consider the construction of
$GSO(-)$ vertices directly in the hybrid formalism, checking the
physical conditions described earlier. In the second part of the
paper we are going to use these vertices to compute the
superstring field theory action, collecting the $GSO(+)$ part in
superfields.

\section{Physical $GSO(-)$  Vertex  Operators in Hybrid Formalism}

In the $GSO(-)$ sector the space time supersymmetry is broken,
however we can still use the set of hybrid variables to explicitly
construct vertex operators. In particular we have built the
physical vertices for the tachyon and the massless fermion by
using only the hybrid variables, showing the appropriate physical
conditions to be satisfied. Most of the calculations are exhibited
explicitly, since they are useful throughout this work.

\subsection{$\Phi^-$ Vertex Operator}
In this subsection we discuss the construction of C-neutral
vertices that are independent of the detailed structure of the
compact manifold. As seen in the previous section, we should have
a tachyon $T$ and the  massless fermions $\chi^-$ in this sector.

\subsubsection{Tachyon Vertex}
There is a unique spin operator with the lowest conformal weight
($W=-\frac{1}{2}$) $\Lambda^{++++}$.  This is the candidate to
represent the part of zero momentum of the tachyon; however, it
must be verified that all physical conditions for this vertex are
indeed satisfied.

Let us consider the  candidate for the tachyon vertex
\begin{eqnarray}
  \Lambda^{++++}T(X)
\end{eqnarray}

We can show that this vertex is fermionic and it is exactly the
zero momentum part of the RNS tachyon vertex in the large Hilbert
space $c \xi e^{-\phi}$. This vertex corresponds to the lowest
level NS sector that is projected out by the GSO projection of
superstring.

Given the vertex above for the Tachyon, it is trivial to verify
that the zero ghost charge and zero C-charge conditions are
satisfied. We still need to check the conditions for the OPEs with
$T$ and $G$.

For simplicity we consider the tachyon vertex with definite
momentum $T(k,z)=\Lambda^{++++}e^{ikX} $ (where $kX$ stands for
$k_m X^m + k^{+i}X^{-i} + k^{-i}X^{+i}$) and we are going to check
the physical conditions for this vertex.

\vspace{0.3cm} \centerline{{\bf $N=2$ Conditions}} \vspace{0.3cm}

Using the stress tensor we get
\begin{eqnarray*}
L(z)T(k,0)&=&\frac{(-\frac{1}{2}+\frac{k^2}{2})}{z^2}T(k,0)
+\frac{1}{z}\partial T(k,0)+\cdots, \\
k^2&=& k_mk^m + k^{+i}k^{-i},
\end{eqnarray*} requiring that the double pole
vanishes we get the on-shell condition for the momentum $k$ of the
tachyon $k^2=1$.

Now we need to verify the $G$ constraints, ensuring the vanishing
of double poles and higher.

We find that the OPE $G^+(z)T(k,w)$ is
\[
\begin{array}{rcl}
(e^{\rho}(d)^2 + \Gamma^{-j}\partial X^{+j})(z)\Lambda^{++++}
e^{ikX}(w)&=& \frac{1}{(z-w)}\{[-2\Lambda^{--++}+k_m (p\sigma^m
\bar
\theta\Lambda^{++++}) \\
&&+\frac{1}{2}(k^mk_m-1)\Lambda^{++\oplus\oplus}]e^\rho \\
&&+ k^{+j}\Gamma^{-j}\Lambda^{++++}]e^{ikX}\} +\cdots
\end{array}
\]

The OPE ${G}^-(z)T(k,w)$ is
\[
\begin{array}{rcl}
(e^{-\rho}(\bar{d})^2 + \Gamma^{+j}\partial
X^{-j})(z)\Lambda^{++++} e^{ikX}(w)&=& \frac{1}{(z-w)}\{[-2
\Lambda^{++--} +
k_m (\bar p \bar \sigma^m \theta \Lambda^{++++}) \\
&&+\frac{1}{2}(k^mk_m+1)\Lambda^{\oplus\oplus--}]e^{-\rho} \\&&+
k^{+j}\Gamma^{-j}\Lambda^{++++}\}e^{ikX} +\cdots
\end{array}
\]
where the notation $(p\sigma\theta \Lambda)$ describes only the
indices structure in the spin field, no contractions implied. We
see immediately that there are no double poles in these OPEs,
which implies that the vertex satisfies the physical conditions
and no polarization conditions are imposed in the momentum $k$. We
have shown that the tachyon vertex $T(k,z)$ is physical.

Now we proceed by showing that the $N=4$ physical conditions are
also satisfied. The only non-trivial part to be shown is that the
equation of motion is indeed satisfied.

\vspace{0.3cm} \centerline{{\bf Equation of Motion }}
\vspace{0.3cm}

The gauge invariant Equation of motion is $[\tilde{G}^+]_0 [G^+]_0
T(k,z)=0$ we first use our result for the OPE in $G^+(z)T(k,w)$
(the notation $[{\cal O}]_nV$ means the residue of the pole of
$(n+d)$ order in the OPE between ${\cal O}$ and $V$, where $d$ is
the conformal weight of the operator $\cal(O)$).

Using the results above, we need to compute the following OPEs
(for the first piece of the operator $\tilde G^+$ the relevant
terms are the cubic poles, for the second piece the relevant terms
are the double poles, this is due to the different $\rho$
dependence in each term):
\begin{eqnarray}
(\bar{d})^2(z)\Lambda^{--++}e^{ikX}(z) &=& \frac{1}{(z-w)^3} (k^m
k_m) \Lambda^{++++}e^{ikX}(w)+\ldots
\nonumber\\
(\bar{d})^2(z)(p\bar \theta \Lambda^{++++})
e^{ikX}(w)&=&\frac{\bar
\sigma^m}{(z-w)^3}k_m\Lambda^{++++}e^{ikX}(w)+\ldots
\nonumber\\
(\bar{d})^2(z)\Lambda^{++\oplus\oplus}
e^{ikX}(w)&=&(-2)\frac{1}{(z-w)^3}\Lambda^{++++}e^{ikX}(w)+\dots,\nonumber\\
\partial X^{-k}(z) k^{+j}e^{ikX} &=& \frac{1}{(z-w)^2} k^{+j}k^{-j}+\dots \nonumber
\end{eqnarray}
with this OPEs we readily see that
\begin{eqnarray}
[\tilde{G}^+]_0[G^+]_0T(k,z) &=& (k^2-1)T(k,z) =0
\end{eqnarray}
which is the correct equation of motion for the tachyon in
momentum representation.

\subsubsection{Massless Fermion Vertices}
On the next level of the spectrum of states, we have two pairs of
spin field vertices with zero conformal weight ($W=0$). It can be
shown that each pair of vertex is bosonic. These vertex are
candidates to be the zero momentum contribution associated with
the massless space-time fermions.

The OPEs between the hybrid variables and hybrid spin fields have
a square root branch cut, from this OPE we may find the excited
spin fields we are looking for. We may consider both $p$ and
$\theta$ (with their conjugate fields) to construct the spin
fields, which indeed have the correct conformal weight:
\begin{eqnarray}
  \label{spinfields}
  \Theta_\alpha &\equiv& :\theta_\alpha \Lambda^{++++} :=
  \spin{\oplus}{+}{+}{\oplus}{+}{+}, \nonumber\\
  \bar \Theta_{\dot\alpha} &\equiv& :\bar\theta_{\dot\alpha} \Lambda^{++++} :=
  \barspin{+}{+}{\oplus}{+}{+}{\oplus}, \nonumber\\
  \Pi_\alpha &\equiv& :p_\alpha \Lambda^{++++}: =
  \spin{-}{+}{+}{-}{+}{+}, \\
  \bar \Pi_{\dot\alpha} &\equiv& :\bar p_{\dot\alpha} \Lambda^{++++} :=
  \barspin{+}{+}{-}{+}{+}{-}.\nonumber
\end{eqnarray}
Using these fields we may construct a fermion vertex for the
massless (4-dimensional) sector:
\begin{equation}
  \label{massfermion4d}
  \Phi=(\Theta_\alpha \chi_{1}^\alpha + \Pi_\alpha \chi_{2}^\alpha +
  \bar\Theta^{\dot\alpha} \bar\chi_{1\dot\alpha} + \bar \Pi^{\dot\alpha}
  \bar \chi_{2\dot\alpha})e^{ikX}
\end{equation}
where we have decided to work in momentum representation and the
$\chi$-fields represent the polarizations of the fermions. We
assume that the momentum $k$ has only components in the four
dimensional space time in order to avoid the mixing with the other
fields. (Remember that we have a fermion in 10-dimensions and its
equation of motion should be a Dirac equation in 10-dimensions,
due to the Lorentz symmetry breaking pattern we are using, the
different fermion vertices in the hybrid formalism will mix with
each other, unless we assume this simplifying condition.)

We are now going to check whether this vertex satisfies the
physical condition or not:

\vspace{0.3cm} \centerline{{\bf N=2 constraints}} \vspace{0.3cm}

The vanishing of the double poles (and higher) in the OPE of the
stress tensor with the vertex operator is quite trivial since only
the first term in eq.(\ref{n=4}) contributes to the double pole,
yielding the mass-shell condition $k^2=0$ if this vertex is to
have zero conformal weight.

The $G$ constraints are a bit more subtle and should give us
polarization conditions. Let us explicitly show the relevant OPEs
and discuss it:
\begin{eqnarray}
  \label{g+massfermion}
  G^+(z)(\Pi \chi_2)(0) &\sim& \frac{e^\rho}{z}\left[ k^2
  \spin{-}{+}{+}{-}{\oplus}{\oplus} \chi_2 + k_m \chi_2 \sigma^m
  \barspin{-}{-}{\oplus}{+}{+}{\oplus} \right]+\cdots \nonumber \\
  G^+(z)(\Theta \chi_1)(0) &\sim& \frac{e^\rho}{z^2}\left[ \Pi \chi_1+ k_m
  \chi_1\sigma^m \bar\Theta \right] + \frac{e^\rho}{z} \left[k^2
  \spin{\oplus}{+}{+}{\oplus}{\oplus}{\oplus} \chi_1 \right] +\cdots
  \nonumber \\
  G^+(z)(\bar\Pi \bar \chi_2)(0) &\sim& \frac{e^\rho}{z^2}\left[ k^2 \bar
  \Theta \bar\chi_2+ k_m \bar\chi_2\bar\sigma^m \Pi \right] +
  \frac{e^\rho}{z} \left[\barspin{-}{-}{-}{+}{+}{-} \bar\chi_2 + k_m
  \bar\chi_2 \sigma^m \spin{\oplus}{+}{+}{\oplus}{\oplus}{\oplus} \right]
  +\cdots \nonumber \\
  G^+(z)(\bar\Theta \bar\chi_1)(0) &\sim& \frac{e^\rho}{z}\left[
  k_m \bar\chi_1\bar\sigma^m \spin{-}{+}{+}{-}{\oplus}{\oplus} +
  \barspin{-}{-}{\oplus}{+}{+}{\oplus}\bar\chi_1 \right]+\cdots
\end{eqnarray}
where we omitted the momentum dependence and contraction in
fermionic indices. The $G^-$ constraints have the same form and
can be read from this expression, changing the $\rho$ dependence
to $e^{-\rho}$ and moving the bars to appropriate places (for
instance $G^-\Pi \chi_2 = \overline{G^+\bar\Pi \bar \chi_2}$).

Vanishing of the double poles in this OPEs imply the following
conditions for the polarizations:
\begin{eqnarray*}
  \chi_1 &=& k_m\sigma^m \bar \chi_2, \\
  \bar \chi_1 &=& k_m\bar \sigma^m \chi_2.
\end{eqnarray*}
A physical vertex is a superconformal primary state defined modulo
spurious states (with zero norm), created by the negative frequency part
of the superconformal generators. So, we need to identify physical states
modulo gauge transformations $\delta \Phi = G^+(\bar\Pi \bar\zeta +\Theta
k_m\sigma^m \bar\zeta) + G^-(\Pi \zeta +\bar \Theta k_m\bar \sigma^m
\zeta)$, since the transformation should be of the form $\delta \Phi =
G\Lambda$, with $\Lambda$ physical. This identification yields the
following gauge transformations for the vertices:
\begin{eqnarray*}
  \delta \bar \chi_1&=& k^2 \bar\zeta\\
  \delta \chi_2 &=& k_m \sigma^m \bar \zeta,
\end{eqnarray*}
and analogous transformations for the other polarizations. So we
realize that not all the spacetime fields in
eq.(\ref{massfermion4d}) are independent. This is the same
situation we encounter in the $GSO(+)$ sector when considering an
$N=1, D=4$ vector superfield as vertex operator for the massless
sector \cite{Berkovits4}. This way we see that we have only two
spacetime massless spinor fields in this sector, with opposite
chiralities.

Now we should check the $N=4$ conditions for this vertex:

\vspace{0.3cm}\centerline{ {\bf $N=4$ Formalism}} \vspace{0.3cm}

The $N=2$ formalism gives us a gauge fixed vertex operator, while
in the $N=4$ formalism we have an equation of motion for the
vertex operator, as well as a gauge invariance.

The equation of motion is obtained by $\tilde G^+ G^+[\Phi]=0$,
where we are going to use the same vertex eq.(\ref{massfermion4d})
as before and $G$ applied to the vertex means the contour
integral, as usual. Since we have already calculated the OPEs
eq.(\ref{g+massfermion}), we may use this result taking care of
the pole structure coming also from the $\rho$-dependence. We may
see that:
\begin{eqnarray}
  \label{eqmotionfermion}
  \tilde G^+ G^+ (\Theta \chi_1) &=& k_m \chi_1 \sigma^m \bar \Pi + k^2 \Theta
  \chi_1, \nonumber \\
  \tilde G^+ G^+ (\bar \Theta \bar \chi_1) &=& k_m\bar \chi_1 \bar \sigma^m
  \Pi + k^2 \bar \Theta  \bar \chi_1, \nonumber \\
  \tilde G^+ G^+ (\Pi \chi_2) &=& k^2 k_m \chi_2 \sigma^m \bar \Theta +
  k^2 \Pi \chi_2, \nonumber \\
  \tilde G^+ G^+ (\bar \Pi \bar \chi_2) &=& k^2 k_m \bar \chi_2 \bar
  \sigma^m \Theta + k^2 \bar \Pi \bar \chi_2.
\end{eqnarray}
and the equation of motion imply:
\begin{eqnarray*}
  k_m \chi_1 \sigma_m + k^2 \bar \chi_2 &=&0,\\
  k_m \bar \chi_1 \bar \sigma_m + k^2 \chi_2 &=&0,\\
  k^2 (k_m \bar\chi_2 \bar \sigma_m + \chi_1) &=&0,\\
  k^2 (k_m \chi_2 \sigma_m + \bar \chi_1) &=&0.
\end{eqnarray*}

The gauge invariances, as described in the section on physical
conditions, are:
\begin{eqnarray*}
  G^+[e^{-\rho}(\Theta \lambda_1 + \bar \Pi \bar \lambda_2)] &=& \Pi
  \lambda_1 + k_m \lambda_1 \sigma^m \bar \Theta + k^2 \bar \Theta\bar
  \lambda_2 + k_m \bar\lambda_2 \bar \sigma^m \Pi,\\
  \tilde G^+[e^{2\rho}\Gamma^{+i}\Gamma^{+j}\Gamma^{+k}(\Pi \lambda_2 +
  \bar \Theta \bar \lambda_1)] &=& \bar \Pi \bar \lambda_1 + k_m \bar
  \lambda_1 \bar \sigma^m \Theta + k^2 \Theta \lambda_2 + k_m \lambda_2
  \sigma^m \bar \Pi.
\end{eqnarray*}
from this conditions we may read the gauge transformations for the
spacetime fields:
\begin{eqnarray*}
  \delta \chi_1 &=& k_m \bar \lambda_1 \bar\sigma^m + k^2 \lambda_2, \\
  \delta \bar \chi_1 &=& k_m \lambda_1 \sigma^m + k^2 \bar\lambda_2, \\
  \delta \chi_2 &=& \lambda_1 +  k_m \bar \lambda_2 \bar\sigma^m, \\
  \delta \bar \chi_2 &=& \bar \lambda_1 + k_m \lambda_2 \sigma^m.
\end{eqnarray*}
All this is in complete analogy with the $GSO(+)$ sector, again we
have only two massless spinor spacetime fields with opposite
chiralities, the usual equation of motion is verified. What
amounts to choose the Wess-Zumino Gauge in the $GSO(+)$ sector is
to choose the spin fields $\Theta$ and $\bar \Theta$ to be in the
vertex operator, this is what we have done in table 1.

The equivalence between the two formalisms described above is
simple, the $N=2$ formalism gives us a gauge-fixed vertex
operator, whose gauge fixing conditions are exactly the $G$
constraints.

\subsection{$\Psi^-$ and $\bar{\Psi}^-$ Vertex  Operators}

In this subsection we construct the vertices that dependent of the
structure of the compact manifold. We represent these vertices as
$\Psi^-$ and $\bar{\Psi}^-$ with C-charge $1/3$ and $-1/3$
respectively. Since these two vertices are conjugate to  each
other we should treat them together.

The candidates for a vertex of this kind are:
\begin{eqnarray*}
\Psi(k,z) &=& e^\rho \Gamma^{+i}\left(\Theta \chi_1 + \Pi \chi_2 +
\bar
  \Theta \bar \eta_1  + \bar \Pi \bar \eta_2\right)e^{ikX}, \\
\bar\Psi(k,z) &=& e^{-\rho} \Gamma^{-i}\left(\Theta \eta_1 + \Pi
\eta_2 + \bar
  \Theta \bar \chi_1  + \bar \Pi \bar \chi_2\right)e^{ikX}.
\end{eqnarray*}

The necessary OPEs were already calculated in
eq.(\ref{g+massfermion}), but the relevant poles are different due
to the contribution of $\rho$ and $\Gamma$. (Note also that we are
still considering the momentum running only in 4 dimensions). Let
 us proceed by checking the physical conditions:

\vspace{0.3cm}\centerline{ {\bf $N=2$ Formalism}} \vspace{0.3cm}

Vanishing of the double pole in the OPE of the stress tensor $L$
and the vertex operator is just like before. Only the first term
and $L$ contributes and the physical condition is the mass-shell
condition $k^2=0$.

In order to check the $G^+$ constraint in the $\Psi$ vertex one
should note that the order of the poles change, but the terms are
maintained in the equation eq.(\ref{g+massfermion}). Besides the
$\rho$ and $\Gamma$ dependence, the poles of this equation acquire
one more negative power. For the $G^-$ constrain we have something
analogous, but the poles acquire one more positive power, since
the higher order poles were second order, and we should ensure the
vanishing of all second and higher order poles, we  see that $G^-$
puts no constraint in $\Psi$, since there are no double poles (or
higher) in this OPE. However, the $G^+$ has additional
constraints:
\begin{eqnarray*}
  \bar \eta_2 &=& 0,\\
  \chi_1 &=& -k_m \bar\eta_2 \bar \sigma^m, \\
  \bar \eta_1 &=& k_m \chi_2 \sigma^m,
\end{eqnarray*}
and we see that there is actually only one massless spacetime
spinor in the vertex operator $\Psi^-$ and since $k^2=0$, $\eta$
satisfies $k_m\bar \eta_1 \sigma^m =0$.

Same reasoning is valid to the $\bar \Psi$ vertex operator, and
only combinations of  $\Theta$ and $\bar \Pi$ survive.

\vspace{0.3cm}\centerline{ {\bf $N=4$ Formalism}} \vspace{0.3cm}

All the necessary computations have already been done in the
$\Phi$ vertex discussed above. In order to avoid repetition, we
just state that we have only two physical massless spinor
spacetime fields in these two vertices and we can choose
(gauge-fixing) $\Psi$ to have only the $\bar \Theta$ combination
and $\bar \Psi$ to have only the $\Theta$ combination, as a result
of the 8 equations of motion and the 8 gauge invariances.

With the vertex operators discussed above we are going to proceed
to the study of the superstring field theory and compute the
contribution of these vertices to it.

\section{Hybrid Superstring Field Theory for non-BPS D-Branes}

The action we are going to use is Berkovits superstring field
theory action \cite{Berkovits7}, which is the only string field
action that allows one to include the Ramond sector in a gauge
invariant way without divergence problems at tree level.

The action constructed in \cite{Berkovits7} is
\begin{eqnarray}
\nonumber    S &=& \langle (e^{-\Phi} G_{-1} e^\Phi)(e^{-\Phi} G_0
e^\Phi)+ (e^{-\Phi}
  G_{\frac{2}{3}} e^\Phi)(e^{-\Phi}
G_{-\frac{1}{3}} e^\Phi) \rangle_D \\
\nonumber &&+ \langle \int_0^1 dt (e^{-\tilde\Phi}\partial_t
e^{\tilde \Phi}) \left( \{e^{-\tilde\Phi}G_{-1}
  e^{\tilde \Phi}, \;e^{-\tilde\Phi}G_0 e^{\tilde \Phi}
\} + \{e^{-\tilde\Phi} G_{-\frac{2}{3}}
  e^{\tilde \Phi}, \;e^{-\tilde\Phi}G_{-\frac{1}{3}}
e^{\tilde \Phi} \}\right)\rangle_D  \\
\nonumber  && -\langle e^{- \Phi} {\bar\Omega} e^{ \Phi} \Omega +
{\bar\Omega} e^{ \Phi}
  G_{-\frac{2}{3}}e^{- \Phi}  +  \Omega e^{-
\Phi}
  G_{-\frac{1}{3}}e^{+ \Phi} \rangle_D \\
  && - \langle \frac{1}{2}  {\bar\Omega}
    G_{-\frac{1}{3}}  {\bar\Omega}+
\frac{1}{3}{\bar\Omega}^3 \rangle_{\bar F}+\langle \frac{1}{2}
{\Omega}
    G_{-\frac{2}{3}}  {\Omega} +
\frac{1}{3}{\Omega}^3 \rangle_{F}
\end{eqnarray}
where the operators $G$ appearing in this action are defined in
eq.(\ref{n=4}), but now with the $+$ superscript dropped in order
to avoid confusion with $GSO$ notation, subscript denotes
C-charge. We should also note that there are three different
correlators. The D-correlator is defined in the large Hilbert
Space of the superstring. The $F$ and $\bar F$ correlators are
chiral (anti-chiral) subspaces defined with the trivial cohomology
pieces of the G operator, $G_{-1}$ and $G_0$.

The string fields appearing in the action are defined in such a
way that $\Phi$ has zero C-charge, $\Omega\equiv G_{-1}\Psi$, is a
chiral field obtained from the string field $\Psi$ with C-charge
$1/3$ and $\bar \Omega \equiv G_0 \bar \Psi$ is an anti-chiral
string field obtained from a string field $\bar \Psi$ with
C-charge $-1/3$. We are using the notation
$\tilde\Phi=t\Phi,\;\;0\leq t=1$.

Also, one should remember that the products between any two string
fields are the Witten's midpoint interaction.  The prescription we
are going to use is:
\begin{eqnarray}
  \langle V_1 V_2 \cdots V_N \rangle
&=& \left( -\frac{4i}{N}\right)^{\sum_{k=1}^N h_k} e^{\frac{2\pi
  i}{N}\sum_{l=1}^N h_l(l-1)} \left\langle \Pi_{j=1}^N V_j\left(
e^{\frac{2\pi i}{N}(j-1)} \right)\right\rangle \nonumber
\end{eqnarray}
where $V_i$ here are conformal primary fields with conformal
weight $h_i$.

The correlator of the conformal field theory fields in the
right-hand side is calculated in the disk, since this prescription
means that we take the $N$ open string world-sheets, represented
by $N$ upper half-disks and map them to the unit disk (defined in
global coordinates) via global conformal transformations, taking
care to glue the right sides of each string. A detailed discussion
of this Witten interaction may be found in \cite{Ohmori1}.

In order to introduce $GSO(-)$ sector, we are going to consider
the following tensor products:
\[
\begin{array}{lcccl}
  \hat \Phi \equiv \Phi^+\otimes I +
\Phi^-\otimes \sigma_1,
 &&&&\\
  \hat \Psi \equiv \Psi^+\otimes I +
\Psi^-\otimes \sigma_1,&&&&
  \hat {\bar\Psi} \equiv \bar\Psi^+\otimes I+ \bar{\Psi}^-\otimes
  \sigma_1, \\
  \hat {\Omega} \equiv  {\Omega}^+\otimes \sigma_3 + {\Omega}^-\otimes
  i{\sigma}_2, &&&&
  \hat {\bar\Omega} \equiv \bar\Omega^+\otimes \sigma_3 +
  \bar\Omega^-\otimes i\sigma_2,
\end{array}
\]
\[
\begin{array}{lcccl}
  \hat G_{0} \equiv  G_{0}\otimes \sigma_3,
  &&&&\hat G_{-1} \equiv  G_{-1}\otimes \sigma_3,\\
  \hat G_{-\frac{1}{3}} \equiv G_{-\frac{1}{3}}\otimes
  \sigma_3,
  &&&&
  \hat G_{-\frac{2}{3}} \equiv G_{-\frac{2}{3}}\otimes \sigma_3
\end{array}
\]
where $\sigma_i$ are Pauli matrices. All this only means that we
split string fields in two pieces according to the GSO sector they
have come out and tensor it by appropriate internal Chan-Paton
factors, since the commutative properties of the two sectors are
different (GSO$(-)$ sector being Grassmann odd), this is to ensure
the ciclicity of the correlators as in any usual CFT. Also the
BRST-like operators need to be tensorized in such a way that, at
the end we get all the properties of a derivation as they act in
the string fields, in particular Leibniz rule should be verified.
It is important to remark only that the (anti)chiral fields
appearing in the action are split in the same way, but as they
have a differential operator acting on it ($G_0$ or $G_{-1}$) we
have the tensor product of $\sigma_3$ carried by the G-operator
and the matrices in the string fields, either $\Psi$ or $\bar
\Psi$. The correlators should include the trace over these
internal Chan-Paton matrices.

\subsection{Vertex Operators in SSFT}

The vertex operator for the Tachyon and the massless sector are:
\[
\begin{array}{lcccl}
  \Phi^+ = v(x,\theta,\bar\theta), &&&&  \Phi^- =
  \Lambda^{++++}T(x) + \Theta^\alpha \chi^-_\alpha(x)
  +\bar\Theta^{\dot\alpha}\bar\chi^-_{\dot\alpha}(x),
   \nonumber \\
  \Psi^+ = e^\rho \Gamma^{+j}  \bar\theta^2\omega_{+j}(x,\theta), &&&&
      \Psi^- = e^\rho \Gamma^{+j} \bar\Theta^{\dot\alpha}
  \bar{\lambda}^-_{+j\dot\alpha}(x), \nonumber\\
  \bar\Psi^+ = e^{-\rho} \Gamma^{-j}
  \theta^2\bar\omega_{-j}(x,\bar\theta),
  &&&&
   \bar\Psi^- = e^{-\rho} \Gamma^{-j}
\Lambda^{\alpha} \lambda^-_{-j\alpha}(x).
\end{array}
\]
These fields are in Table 1. Here we have collected only the
$GSO(+)$ string fields in $D=4, N=1$ superfields, whose component
expansion are:
\[
\begin{array}{rcl}
v(x,\theta,\bar\theta) &=& \theta\sigma^m \bar\theta A_m(x) +
  \theta\theta\bar\theta \bar{\chi}^+(x) +
\bar\theta\bar\theta\theta
  \chi^+(x) + \theta\theta\bar\theta\bar\theta
D(x),\\
\omega_{-i}(x,\theta)&=& A_{-i}(x) +\theta \lambda^+_{-i}(x)
+\theta\theta
  F_{-i}(x),\\
\bar\omega_{+i}(x,\bar\theta)&=& A_{+i}(x) +\bar\theta
  \bar{\lambda}^+_{+i}(x)
  +\bar\theta\bar\theta \bar F_{+i}(x).
\end{array}
\]
The $GSO(-)$ sector is directly taken from the Table 1 and written
in the appropriate string fields, according to their C-charge.

We also need to calculate the (anti)chiral superstring fields that
appear in the action. The idea is to pick the simple pole in the
OPE of the integrand of the appropriate $G$ with the string
fields.
\begin{eqnarray}
  \Omega^+ &=& G_{-1} \Psi^+ = \oint dw \left(
\frac{1}{12} \bar d^{\dot\alpha} \bar d_{\dot\alpha} e^{-2\rho}
  \epsilon^{jkl}
\Gamma^{-j}\Gamma^{-k}\Gamma^{-l}(w)\right) \left( e^\rho
\Gamma^{+j} \omega^{-j}(x,\theta)
  \bar\theta^2\right(z))  \nonumber \\
&=& \oint dw\frac{1}{12} \frac{2}{(w-z)^2} (w-z)^2e^{-\rho}
6\epsilon^{jkl} \Gamma^{-j} \Gamma^{-k} \frac{1}{z}
  \omega^{-l}           \nonumber \\
&=& e^{-\rho} \Gamma^{-j}\Gamma^{-k} \omega^{-l}\epsilon^{jkl}.\\
&&\nonumber\\
  \bar\Omega^+ &=& G_{0} \bar\Psi^+ = \oint dz
\left(\frac{1}{2} d^\alpha d_\alpha e^\rho (w)\right)
  \left(e^{-\rho} \Gamma^{-j}
\bar\omega^{+j}(x,\bar\theta) \theta^2 (z)\right)
\nonumber \\
  &=& \oint dw \frac{1}{2} \frac{2}{(w-z)^2} z
\Gamma^{-j} \bar \omega^{+j} \nonumber\\
  &=& \Gamma^{-j} \bar \omega^{+j}.\\
&&\nonumber\\
  \Omega^- &=& G_{-1}\Psi^- = \oint dw\left(
\frac{1}{12}\bar d^{\dot\alpha} \bar d_{\dot\alpha} e^{-2\rho}
  \epsilon^{jkl}
\Gamma^{-j}\Gamma^{-k}\Gamma^{-l}(w)\right)\left( e^\rho
\Gamma^{+i} \bar \Theta \bar \lambda^{-i}(z)
  \right)  \nonumber \\
  &=& \oint dw\frac{1}{12} \left( (w-z)^2 e^{-\rho}
\right) \left(
\frac{6\epsilon^{jkl}\Gamma^{-j}\Gamma^{-k}\eta^{il}}{z}\right)
\left(
  \frac{2\bar\Pi\bar\lambda^{-i}+i\partial_m
\bar\lambda^{-j}\bar\sigma^m\Theta}{z^2}
\right) \nonumber\\
  &=& e^{-\rho}
\epsilon^{jkl}\Gamma^{-j}\Gamma^{-k}\left(\bar\Pi\bar\lambda^{-i}+\frac{i}{2}\partial_m
  \bar\lambda^{-j}\bar\sigma^m\Theta \right).
\\
&&\nonumber\\
  \bar\Omega^- &=& G_0 \bar\Psi^- =\Gamma^{-j}
\left(\Pi\lambda^{-i}+\frac{i}{2}\partial_m
  \lambda^{+j}\sigma^m\bar\Theta  \right).
\end{eqnarray}

\section{Tachyon and Massless Sector SSFT
  Action}

We are now almost ready to calculate the superstring field theory
action for the tachyon and massless sector, the only piece of
information still lacking is the definition of the non-vanishing
norms, i.e., a choice for the normalization of the correlators:
\begin{eqnarray}
  \langle \frac{1}{24} \theta^2 {\bar\theta}^2
e^{-\rho} \epsilon^{ijk} \Gamma^{-i}\Gamma^{-j}\Gamma^{-k}
  \rangle_D&=&1,                      \\
  \langle G_0\left(\frac{1}{24} \theta^2
{\bar\theta}^2 e^{-\rho} \epsilon^{ijk}
\Gamma^{-i}\Gamma^{-j}\Gamma^{-k}
  \right)\rangle_{\bar F} &=& \langle
\frac{1}{24} {\bar\theta}^2 \epsilon^{ijk}
 \Gamma^{-i}\Gamma^{-j}\Gamma^{-k}\rangle_{\bar
F} = 1,                 \\
 \langle G_{-1}\left(\frac{1}{24} \theta^2
{\bar\theta}^2 e^{-\rho} \epsilon^{ijk}
  \Gamma^{-i}\Gamma^{-j}\Gamma^{-k}
\right)\rangle_{ F} &=& \langle\frac{1}{4}e^{-3\rho} \theta^2
(\Gamma \cdot  \partial\Gamma)^3\rangle =1.
\end{eqnarray}
The first norm is exactly the non-vanishing norm in the large
Hilbert Space of the Superstring, in RNS variables this is $\xi
e^{-2\phi} c\partial c\partial^2 c$, giving us the correct
background charge for all the ghost fields.  Its value is related
with the superconformal killing vectors of the disk as discussed
in \cite{Polchinski}.  The non-vanishing norms in the (anti)chiral
subspaces are obtained from the norm of the large Hilbert space by
the application of the suitable differential operator (with
trivial cohomology) which defines the (anti)chiral subspace.

The Superstring Field Theory Action is non-polynomial, in order to
deal with it we are going to expand the exponentials and keep
terms up to four string fields. For the vertices we are
considering, all other terms vanish since it is impossible to
write higher order terms with the string fields at first and
second mass levels that cancel the background charge yielding
non-vanishing correlators.

For the first part of the action we have:
\begin{eqnarray}
\label{swz1} S^{(1)}_{WZ}&=&\langle \left(e^{-\hat\Phi}\hat
G_{-1}e^{\hat\Phi}\right)\left(e^{-\hat\Phi}\hat
  G_{0}e^{\hat\Phi}\right) \rangle_D - \langle
  \int_0^1dt(e^{-\tilde {\hat\Phi}}\hat
\partial_{\tau}e^{\tilde
  {\hat\Phi}})\{e^{-\tilde{\hat\Phi}}G_{-1}
  e^{\tilde{\hat \Phi}}, e^{-\tilde{\hat\Phi}}G_0 e^{\tilde{\hat
  \Phi}}
\}\rangle_D \nonumber \\
  &=& 2\sum_{M,N=0}^{\infty}
\frac{(-1)^N}{(M+N+2)!}
    \begin{pmatrix}
      M+N \\ N
    \end{pmatrix}
   \langle \left( \hat{G}_{-1}
\hat\Phi\right)\hat\Phi^M \left(
   \hat{G}_{0} \hat\Phi\right)\hat\Phi^N
\rangle_D                                    \nonumber \\
  &=& 2 \left[ \frac{1}{2} \langle \left(
\hat{G}_{-1} \hat\Phi\right)
   \left( \hat{G}_{0} \hat\Phi\right) \rangle_D -
\frac{1}{6} \langle \hat\Phi
   \left\{ \hat{G}_{-1} \hat\Phi\right\}
\left\{ \hat{G}_{0} \hat\Phi\right\}
   \rangle_D \nonumber \right.\\&&\left. - \frac{1}{24}\langle \left[
\hat\Phi, \hat{G}_{-1} \hat\Phi\right] \left[\hat\Phi,
   \hat{G}_{0} \hat\Phi\right] \rangle_D+\cdots
\right]
\nonumber \\
  &=& \langle \left( \hat{G}_{-1}
\hat\Phi\right)
   \left( \hat{G}_{0} \hat\Phi\right) \rangle_D -
\frac{1}{3}\langle  \hat\Phi
   \left\{ \hat{G}_{-1} \hat\Phi\right\}
\left\{ \hat{G}_{0} \hat\Phi\right\}
   \rangle_D - \frac{1}{12}\langle \left[
\hat\Phi, \hat{G}_{-1} \hat\Phi\right] \left[\hat\Phi,
   \hat{G}_{0} \hat\Phi\right] \rangle_D
\nonumber\\ &&+\cdots 
\end{eqnarray}

Splitting in $GSO(+)$ and $GSO(-)$ fields and taking care of the
Pauli matrices multiplying the different string fields and
operators, the relevant terms we need to compute are the
following:
\begin{eqnarray}
\langle \hat{G}_{-1} \hat\Phi, \hat{G}_{0} \hat\Phi \rangle_D &=&
\underbrace{\langle G_{-1} \Phi^+ G_{0} \Phi^+\rangle_D}_{GSO(+)}
-\langle G_{-1} \Phi^-G_{0} \Phi^- \rangle_D,
\end{eqnarray}
\begin{eqnarray}
\langle -\frac{1}{3} \hat\Phi \left\{ \hat{G}_{-1} \hat\Phi,
\hat{G}_{0}
  \hat\Phi \right\}\rangle_D &=&
\underbrace{\langle \frac{1}{3} \Phi^+ \left\{ G_{-1} \Phi^+, G_0
  \Phi^+ \right\}\rangle_D}_{GSO(+)}
  +\langle
\frac{1}{3} \Phi^+ \left\{ \hat{G}_{-1} \Phi^-, \hat{G}_{0}
  \Phi^- \right\}\rangle_D \nonumber \\ && -
\langle \frac{1}{3} \Phi^- \left[ G_{-1} \Phi^+, G_{0}
  \Phi^- \right] \rangle_D+\nonumber \\&&+ \langle \frac{1}{3}
\Phi^- \left[ G_{-1} \Phi^-, G_{0}
  \Phi^+ \right] \rangle_D,
\end{eqnarray}
\begin{eqnarray}
\langle -\frac{1}{12} \left[ \hat\Phi, \hat{G}_{-1}
\hat\Phi\right] \left[\hat\Phi,
   \hat{G}_{0} \hat\Phi\right] \rangle_D
&=&\underbrace{ -\langle
   \frac{1}{12} \left[ \Phi^+, G_{-1}
\Phi^+\right]
   \left[\Phi^+, G_{0} \Phi^+\right]
\rangle_D}_{GSO(+)}\nonumber \\
 && - \langle
   \frac{1}{12} \left\{ \Phi^-, G_{-1}
\Phi^-\right\}
   \left\{\Phi^-, G_{0} \Phi ^-\right\}
\rangle_D \nonumber \\ &&
  + \langle \frac{1}{12} \left[ \Phi^+,
G_{-1} \Phi^+ \right] \left\{\Phi^-,
   G_{0} \Phi^-\right\} \rangle_D \nonumber
\\ &&
  +\langle \frac{1}{12} \left\{ \Phi^-,
G_{-1} \Phi^-\right\} \left[\Phi^+,
   G_{0} \Phi^+\right] \rangle_D \nonumber
\\ &&
  -\langle \frac{1}{12} \left[ \Phi^+,
G_{-1} \Phi^-\right] \left[\Phi^+,
   G_{0} \Phi^-\right] \rangle_D \nonumber
\\ &&
  -\langle \frac{1}{12} \left\{ \Phi^-,
G_{-1} \Phi^+\right\}
   \left\{\Phi^-, G_{0} \Phi^+ \right\}
\rangle_D \nonumber \\ &&
  + \langle \frac{1}{12} \left[ \Phi^+,
G_{-1} \Phi^-\right] \left\{\Phi^-,
   G_{0} \Phi^+\right\} \rangle_D
\nonumber \\ &&
  +\langle \frac{1}{12} \left\{ \Phi^-,
G_{-1} \Phi^+\right\} \left[\Phi^+,
   G_{0} \Phi^-\right] \rangle_D.
\end{eqnarray}
Note that only even number of $GSO(-)$ string fields make no
vanishing contribution.

Now we need to compute these correlators. Most of the calculations
are tedious but straightforward. We consider all the spin fields
bosonized, which allow us to implement a routine to calculate the
correlators in programs of symbolic computation to get the correct
factors. Since bosonization breaks Lorentz invariance, extra care
need to be taken in order to write the results in Lorentz
invariant manner.
\begin{eqnarray}
  \langle -G_{-1} \Phi^- G_0 \Phi^- \rangle_D
&=& T \Box T +\frac{1}{2} T^2 +2 i
  \left( \chi^- \sigma^m \partial_m
\bar{\chi}^- + \bar{\chi}^- \bar\sigma^m \partial_m \chi^-
\right),
\nonumber \\
&&                \\
  \langle -\frac{1}{3} \hat\Phi \left\{
\hat{G}_{-1} \hat\Phi, \hat{G}_0
  \hat\Phi \right\} \rangle_D &=&
  4\  3^{3/4} T
\left( \chi^+ \chi^- +
  \bar{\chi}^+ \bar{\chi}^- \right)   \nonumber
\\
  &&
+ 2 \left( \chi^+ \sigma^m \left[
  A_m,\bar{\chi}^- \right]
 + \bar{\chi}^-
  \bar\sigma^m \left[ A_m,\chi^+ \right]
\right), \\
  \langle -\frac{1}{12} \left\{ \Phi^-, G_{-1}
\Phi^- \right\} \left\{
  \Phi^-, G_0 \Phi^- \right\} \rangle_D &=&
-T^4,
\\
  \langle \frac{1}{12} \left[ \Phi^+, G_{-1}
\Phi^+ \right] \left\{
  \Phi^-, G_0 \Phi^- \right\} \rangle_D &=& 0,
\\
 \langle -\frac{1}{12} \left\{ \Phi^-, G_{-1}
\Phi^- \right\} \left[
  \Phi^+, G_0 \Phi^+ \right] \rangle_D &=& 0,
\\
  \langle -\frac{1}{12} \left[ \Phi^+, G_{-1}
\Phi^- \right] \left[
  \Phi^+, G_0 \Phi^- \right] \rangle_D &=&
\frac{1}{2} [T,A_m][T,A^m],
\end{eqnarray}
\begin{eqnarray}
  \langle \frac{1}{12} \left\{ \Phi^+, G_{-1}
\Phi^- \right\}  \left\{
  \Phi^-, G_0 \Phi^+ \right\} \rangle_D &=& 0,
\\
  \langle -\frac{1}{12} \left\{ \Phi^-, G_{-1}
\Phi^+ \right\} \left\{
  \Phi^-, G_0 \Phi^+ \right\} \rangle_D &=&
\frac{1}{2} [T,A_m][T,A^m],
\\
  \langle \frac{1}{12} \left\{ \Phi^-, G_{-1}
\Phi^+ \right\} \left[
  \Phi^+, G_0 \Phi^- \right] \rangle_D &=& 0.
\end{eqnarray}
For the second Wess-Zumino part of the action things are easier,
since most of the terms vanish. Only the following term
contributes
\begin{eqnarray}
\label{swz2}
  S^{(2)}_{WZ} &=& -\langle G_{-2/3} \Phi^-
G_{-1/3} \Phi^- \rangle_D =
  4\partial_{-i} T \partial_{+i} T.
\end{eqnarray}
There is still some terms in the D correlator:
\begin{eqnarray}
\label{sd}
  S_D &=& - \langle e^{-\hat\Phi}
\hat{\bar\Omega} e^{\hat\Phi}
  \hat{\Omega}\rangle_D + \langle
\hat{\bar\Omega} e^{\hat\Phi} \hat{G}_{-2/3}
  e^{-\hat\Phi} \rangle_D + \langle
\hat{\Omega} e^{-\hat\Phi} \hat{G}_{-1/3}
  e^{\hat\Phi} \rangle_D.
\end{eqnarray}
Splitting these terms in $GSO(+)$ and $GSO(-)$ fields as before
\begin{eqnarray}
  \langle -e^{-\hat\Phi} \hat{\bar\Omega}
e^{\hat\Phi}
  \hat{\Omega}\rangle_D &=&
\underbrace{\langle \bar\Omega^+
  \Omega^+\rangle_D}_{GSO(+)}
  + \langle
\bar\Omega^-
  \Omega^-\rangle_D            \nonumber\\
  &&
  - \langle \left[
\Phi^+,\bar\Omega^- \right] \Omega^- \rangle_D
 - \langle\left[ \Phi^-,\bar\Omega^- \right]
\Omega^+ \rangle_D \nonumber
  \\ &&+ \langle
  \left[ \Phi^-,\bar\Omega^+ \right] \Omega^-
\rangle_D                   \nonumber \\
&&-\frac{1}{2} \langle \left\{\Phi^-
  \left\{ \Phi^-,\bar\Omega^+ \right\} \right\}
\Omega^+  \rangle_D,
\end{eqnarray}
\begin{eqnarray}
  \langle \hat{\bar\Omega} e^{\hat\Phi}
\hat{G}_{-2/3}
  e^{-\hat\Phi} \rangle_D &=& - \langle
\bar\Omega^+ \left\{ \Phi^-,
  G_{-2/3} \Phi^- \right\}\rangle_D,
\end{eqnarray}
\begin{eqnarray}
  \langle \hat{\Omega} e^{-\hat\Phi}
\hat{G}_{-1/3}
  e^{\hat\Phi} \rangle_D &=& \langle \Omega^+
\left\{ \Phi^-,
  G_{-1/3} \Phi^- \right\}\rangle_D.
\end{eqnarray}
The correlators are calculated in the same way as before, giving
us the mixing of the 4-dimensional and the internal 6-dimensional
fields.
\begin{eqnarray}
  \langle \bar\Omega^- \Omega^-\rangle_D &=& -
i \left( \lambda^-_{-i}
  \sigma^m \partial_m \bar\lambda^-_{-i} +
\bar\lambda^-_{-i} \sigma^m
  \partial_m \lambda^-_{+i} \right),
\\
  \langle -\left[ \Phi^+,\bar\Omega^- \right]
\Omega^- \rangle_D &=& -2
  \left[ A_m, \lambda^-_{+i}\right] \sigma^m
\lambda^-_{+i},
\\
  \langle -\left[ \Phi^-,\bar\Omega^- \right]
\Omega^+ \rangle_D &=& 4
  \lambda^-_{+i} \chi^- A_{+i} - 2\  3^{3/4}
T \lambda^+_{+i} \lambda^-_{-i},
\\
   \langle\left[ \Phi^-,\bar\Omega^+ \right]
\Omega^- \rangle_D &=& 4
 \bar{\lambda}^-_{-j} \bar{\chi}^- A_{-j} - 2\
3^{3/4} T \bar{\lambda}^-_{-i} \bar{\lambda}^+_{+i},
\\
  \langle -\frac{1}{2}\left\{\Phi^-
  \left\{ \Phi^-,\bar\Omega^+ \right\} \right\}
\Omega^+  \rangle_D &=& 2\
  T^2 A_{+i} A_{-i},
\\
   \langle -\bar\Omega^+ \left\{
\Phi^-,G_{-2/3} \Phi^- \right\}\rangle_D
   &=& 4\sqrt{3} A^{+i} T \partial_{+i} T,
\\
  \langle \Omega^+ \left\{ \Phi^-,G_{-1/3}
\Phi^- \right\}\rangle_D &=& 4\sqrt{3} A^{-i} T
\partial_{-i} T.
\end{eqnarray}
and finally the two Chern-Simons terms contribute with:
\begin{eqnarray}
\label{scs2}
  \bar{S}_{CS} &=& -\langle \frac{1}{2}
\hat{\bar\Omega} \hat{G}_{-1/3}
  \hat{\bar\Omega}\rangle_{\bar{F}} - \langle
\frac{1}{3} \hat{\bar\Omega}
  \hat{\bar\Omega} \hat{\bar\Omega}
\rangle_{\bar{F}}
  \nonumber \\
  &=&
 \underbrace{-\langle \frac{1}{2}
\bar\Omega^+ G_{-1/3}
  \bar\Omega^+ \rangle_{\bar{F}}
  - \langle
\frac{1}{3} \bar\Omega^+
  \bar\Omega^+  \bar\Omega^+
\rangle_{\bar{F}}}_{GSO(+)}
  \nonumber \\
&&
  -\langle \frac{1}{2}
\bar\Omega^- G_{-1/3}
  \bar\Omega^- \rangle_{\bar{F}} - \langle
\frac{1}{3} \bar\Omega^+
  \bar\Omega^-  \bar\Omega^- \rangle_{\bar{F}}
  \nonumber, \\
&=&  - \frac{1}{2} \epsilon_{ijk} \lambda^+_{+i} \partial_{-j}
  \lambda^+_{+k} - \frac{1}{3} \epsilon_{ijk}
A_{+i} \lambda^+_{+j} \lambda^+_{+k}
 \nonumber \\
&&  - \frac{1}{2} \epsilon_{ijk} \lambda^-_{+i}
\partial_{-j}
  \lambda^-_{+k} - \frac{1}{3} \epsilon_{ijk}
A_{+i} \lambda^-_{+j} \lambda^-_{+k}.
\end{eqnarray}
\begin{eqnarray}
\label{scs1}
   S_{CS} &=& \langle \frac{1}{2} \hat\Omega
\hat{G}_{-2/3}
  \hat\Omega\rangle_F + \langle \frac{1}{3}
\hat{\Omega}
  \hat{\Omega} \hat{\Omega} \rangle_F
  \nonumber \\
  & =& \frac{1}{2} \epsilon_{ijk}
\bar\lambda^+_{-i} \partial_{+j}
  \bar\lambda^+_{-k} + \frac{1}{3} \epsilon_{ijk}
A_{-i} \bar{\lambda}^+_{-j} \bar{\lambda}^+_{-k}           \nonumber\\
&&+\frac{1}{2} \epsilon_{ijk} \bar{\lambda}^-_{-i} \partial_{+j}
  \bar{\lambda}^-_{-k} + \frac{1}{3} \epsilon_{ijk}
A_{-i} \bar{\lambda}^-_{-j} \bar{\lambda}^-_{-k} .
\end{eqnarray}

The action we got after all this computation is simply:

\begin{equation*}
  \begin{array}{ccccccccccc}
  S &=& S^{(1)}_{WZ}&+&S^{(2)}_{WZ}&+&S_D&+&S_{CS}&+&\bar S_{CS}\\
  && \textrm{eq.(\ref{swz1})} && \textrm{eq.(\ref{swz2})} && \textrm{eq.(\ref{sd})} &&
  \textrm{eq.(\ref{scs1})} &&\textrm{eq.(\ref{scs2})}.
\end{array}
\end{equation*}

Now we may rewrite the 4-dimensional action obtained from all
these terms using 10-dimensional fields. In order to do this we
should take into account the pattern of Lorentz symmetry breaking
eq.(\ref{lorbreak}) and read, from the expressions calculated
above, how they fit in explicit 10-dimensional Lorentz invariance.

The action is \footnote{See footnote in the introduction}
\begin{eqnarray}
  \nonumber S&=& \mathrm{Tr} \int d^{10} x
\left[ \frac{1}{4} F^{\mu\nu}F_{\mu\nu}+ T\Box T +\frac{1}{2}T^2
  +[T,A^{\mu}][T,A_{\mu}]-T^4
\right. \nonumber \\
&& + \chi^{a\;+} \gamma^{\mu}_{ab} \left(\partial_\mu
\chi^{b\;+}+[A_{\mu},\chi^{b\;+}] \right) + \chi^{a\;-}
\gamma^{\mu}_{ab} \left(\partial_\mu
\chi^{b\;-}+[A_{\mu},\chi^{b\;-}] \right)
\nonumber\\
&&\left. + 3^\frac{3}{4} T (\chi^{a\;+} \chi^-_{a})+\cdots \right]
\end{eqnarray}

\section{Summary and Discussion}

On the first part of this paper we have developed the general
framework for the computation of vertices with negative $GSO$
projection in the context of the hybrid formalism. To do that we
have identified this sector with the antiperiodic sector of the
hybrid variables. In general, all $GSO(-)$ vertex operator
requires a hybrid spin field. This hybrid spin field turns out to
be the part of zero momentum of the tachyon $\Lambda^{++++}$.

Using this framework we have explicitly identified some vertex
operators for the $GSO(-)$ sector of the superstring. In
particular, we have computed the vertices for tachyon and massless
fermions in both sectors (Ramond sector). We did the calculation
in the hybrid formalism, where the fields are split in an explicit
4-dimensional part and an internal 6-dimensional part. Again
reminding that internal for our present purposed, only mean that
these are fields depending on the 6 complexified coordinates.

These vertices are useful to understand supersymmetric related
issues in superstring field theory, specially in what regards to
SUSY breaking in tachyon condensation processes.

On the second part of this paper, after computing the action for
the string field theory including up to massless terms in both GSO
sectors (which amount to be a type IIA on a non-BPS D-brane) we
arrive in a full Lorentz covariant action in 10-D, including the
Ramond sector. We note that our action now include the massless
fermions of the action, which was not found in the previous
literature.

Yoneya \cite{Yoneya2} argue that the full string field theory
action including both GSO projections should have a non-linear
supersymmetry since when the system undergoes Tachyon condensation
we recover a closed string vacuum. This non-linear SUSY should
transform fields of all masses - infinite scalars acquire vacuum
expectation values in the closed string vacuum, but the existence
of the tri-linear term in our action shows that we may, at least
naively, consider the arguments in \cite{Yoneya2} for non-linear
SUSY realizations, by considering the $GSO(-)$ fermions as
goldstinos with a non-linear term in its SUSY transformation that
can be calculated order by order. Of course this approach is
hopeless in the current presentation of the problem since the full
action involves an infinite number of terms, however, this
questions are still under work.

One may try to consider some other possibilities, including the
vacuum superstring field theory in the presence of fermions or
Boundary String Field Theory. After the present first step, these
issues are under investigation. Any other approach used may rely
on this calculation by considering the suitable field
redefinitions.

\section{Acknowledgements}
We thank Nathan Berkovits for useful discussions at various stages
of this work. LB would also like to thank Brenno Vallilo and
Amilcar Rabelo de Queiroz for helpful discussions. LB is supported
by FAPESP grant 98/00452-7. CT thanks the Instituto de Fisica
Teorica at Sao Paulo for the hospitality during the preparation of
this work, and FAPESP for the financial support under grant
96/12280-0.

\begin{appendix}
\section{RNS variables}
$N=1,\; c=15$ superconformal generators $T_m =-\frac{1}{2}\partial
x^{\mu}\partial x_{\mu} - \frac{1}{2}{\psi}^{\mu}\partial
{\psi}_{\mu},\;\;
G_m = i\psi^{\mu}\partial x_{\mu}.$\\
BRST operator
\begin{equation*}
    Q=\int\left[c(T_m-b\partial
c-\partial^2\phi-\frac{1}{2}(\partial\phi)^2
    -\eta\partial\xi)+\eta
    e^{\phi}G_m-\eta\partial\eta
e^{2\phi}b\right].
\end{equation*}
Picture Raising: $ Z=e^{\phi}G_m + b\partial\eta
e^{2\phi}+\partial (b\eta e^{2\phi})
+c\partial\xi$.\\
Picture Lowering: $ Y=c\partial \xi e^{-2\phi}.
\\
$C-Charge: $ C= P+\frac{i}{3}\int dz(\psi^4\psi^5 + \psi^6\psi^7
+\psi^8\psi^9). $\\
Picture Number: $
P=\int dz (\xi\eta-\partial\phi).\\
$ Ghost Charge: $ J_{gh} = \int dz ( \eta\xi + cb
). $\\
\centerline{Table 2: Properties of the Extended RNS Variables}
\begin{tabular}{|c|c|c|c|c|c|c|c|c|c|c|c|c|}
\hline
          & $b$ & $c$ & $\xi$ & $\eta$
&$e^{q\phi}$ & $x^m$ & $\psi^m$ & $\psi^{+j}$ &
$\psi^{-j}$ & $e^{qH}$ & $x^{+j}$ &  $x^{-j}$\\
\hline
 $C$ & $0$ & $0$ &
$1 $ & $-1$ & $q$ & $0$ & $ 0 $ & $\frac{1}{3}$ & $-\frac{1}{3}$ &
$q$ &  $0$ & $0$ \\
\hline
 $P$ &   $0$ & $0$ & $1$ & $-1$ & $q$ & $0$ &
$0$ &
 $0$
 & $0$& $0$ &  $0$ & $0$\\
\hline
 $J_{gh}$ &  $-1$ & $1$ & $-1$ & $1$ & $0$ &$0$
& $0$ &
 $0$
 & $0$& $0$ &  $0$ & $0$\\
 \hline
 $U(1)$ &  $0$ &$0$ &$0$& $0$ & $0$ & $0$ & $0$
&
$1$ & $-1$ & $q$ &  $0$ & $0$\\
\hline
 $W$ &  $2$ & $-1$ &
 $0$ & $1$
 & $ -\frac{1}{2}q(q+2)$ & $0$ & $\frac{1}{2}$
& $\frac{1}{2}$ &
 $ \frac{1}{2}$ & $ \frac{3}{2}q^2 $ &  $0$ &
$0$\\
 \hline
 Statistic & F & F & F & F  &  & B & F  & F & F
 & & B & B
\\
  \hline
\end{tabular}\\
 Non-vanishing inner product
\[
\langle \xi e^{-2\phi}c\partial c\partial^2 c \rangle =1.
\]
Hermiticity Relation $j_{BRST}=b^{\dagger}$
\begin{eqnarray*}
j_{BRST} &=& e^{-K} (-b\eta\partial\eta
e^{2\phi})e^{K}, \nonumber \\
 K &=& \int [ c\xi e^{-\phi}\psi^{\mu}\partial
x_{\mu}
 +\frac{1}{2}\partial\phi c\partial c
\xi\partial\xi e^{-2\phi}].
\end{eqnarray*}

Bosonization of $\psi^m$
\begin{eqnarray*}
\psi^{\pm }_1 \equiv e^{\pm \tau_{1}}  &\cong &
 \frac{1}{\sqrt{2}}(\psi^3\pm \psi^0),
\nonumber\\
\psi^{\pm}_2 \equiv e^{\pm\tau_{2}}  &\cong &
\frac{1}{\sqrt{2}}(\psi^1\pm i\psi^2).
\end{eqnarray*}

Bosonization of $\psi^{\pm j}$
\begin{eqnarray*}
 \psi^{\pm j}\equiv e^{\pm\tau_{j+2}} & \cong &
\frac{1}{\sqrt{2}}(\psi^{2j+2}\pm i\psi^{2j+3});\;\;\; j=1,2,3.
\end{eqnarray*}

Four dimensional RNS spin fields $\Sigma^{\alpha}$ and
$\bar{\Sigma}^{\dot\alpha}$
\begin{eqnarray}
\Sigma^{\alpha} \equiv \Sigma^{\pm\pm}   & \equiv &
e^{\pm\frac{1}{2}\tau_1 \pm\frac{1}{2}\tau_2}
,\;\;\;(even \;+),  \nonumber \\
\bar{\Sigma}^{\dot\alpha} \equiv \Sigma^{\pm\pm}   & \equiv &
e^{\pm\frac{1}{2}\tau_1 \pm\frac{1}{2}\tau_2}, \;\;\;(odd \;+).
\end{eqnarray}
Four-dimensional spinor index
\[
\begin{array}{|c|c|}
\hline
    \alpha & \dot\alpha \\
    \hline
 1\equiv ++  & \dot 1 \equiv +- \\
 2\equiv --  & \dot 2 \equiv -+ \\
 \hline
\end{array}
\]
\\
Internal RNS Spin Fields: $\Xi^{a}$ and $\bar{\Xi}^{\dot a}$
\begin{eqnarray}
\Xi^{a} \equiv \Xi^{\pm\pm\pm}   & \equiv &
e^{\pm\frac{1}{2}\tau_3 \pm\frac{1}{2}\tau_4
                   \pm\frac{1}{2}\tau_5
}, \;\;\;(even \;+),                 \nonumber\\
\bar{\Xi}^{\dot {a}} \equiv \Xi^{\pm\pm\pm}   & \equiv &
e^{\pm\frac{1}{2}\tau_3 \pm\frac{1}{2}\tau_4
        \pm\frac{1}{2}\tau_5    }, \;\;\;(odd
\;+).
\end{eqnarray}
Internal spinor indices $ a$ (chiral), $\dot{a}$ (anti-chiral)
\begin{equation}
\begin{array}{|c|c||c|c|}
\hline \multicolumn{2}{|c|}{\Xi^a} &
\multicolumn{2}{|c|}{\Xi^{\dot{a}}}\\
\hline
 e^{H/2} & \Xi^{---} & e^{-H/2} & \Xi^{+++} \\
\hline
          & \Xi^{-++} &          & \Xi^{+--} \\
 \Xi^{+j} & \Xi^{+-+} & \Xi^{-j} & \Xi^{-+-} \\
          & \Xi^{++-} &          & \Xi^{--+}\\
\hline
\end{array}    \nonumber
\end{equation}
\section{Hybrid variables}
C-Charge: $ C=\frac{1}{3}\int dz
\Gamma^{+j}\Gamma^{-j}. $ \\
Ghost Charge $ J_{gh} = \int dz (\partial \rho +
\Gamma^{-j}\Gamma^{+j}  ).$
\\
\centerline{Table 3: Properties of the Hybrid Variables }
\begin{tabular}{|c|c|c|c|c|c|c|c|c|c|c|c|}
\hline
   &$X^m$   & $p$ & $\theta$ &
           $\bar{p}$ & $\bar{\theta}$ &
$e^{q\rho}$  & $\Gamma^{+j}$ & $\Gamma^{-j}$ & $X^{+j}$  &
$X^{+j}$ \\
\hline
 $C$  & 0 &$0$ & $0$ & $0$ & $0$ & $0$ &
$\frac{1}{3}$ &
 $-\frac{1}{3}$ & 0 & 0\\
\hline
 $P$    & 0 & $-\frac{1}{2}$ & $\frac{1}{2}$ &
$\frac{1}{2}$ & $-\frac{1}{2}$  &$q$ &  $0$ & $0$
& 0 & 0\\
\hline
 $J_{gh}$ & 0 & $0$   & $0$  & $0$ &$0$   & $q$
& $-1$ &
 $1$ & 0 & 0\\
\hline
 $\rho$ & 0 & $0$ & $0$ &  $0$ & $0$  & $q$  &
$0$ & $0$ & 0 & 0\\
\hline
 $W$ & 0 & $1$  & $0$  & $1$ & $0$  &
$-\frac{1}{2}q(q+1)$ &  $1$ &
 $0$ & 0 &0 \\
\hline
 Statistic & B & F & F & F & F & & F & F & B &
B \\
 \hline
\end{tabular}
\\
 Non-vanishing inner product
\[
\langle \frac{1}{24} (\theta)^2 (\bar\theta)^2
e^{-\rho}\epsilon^{jkl}\Gamma^{-j}\Gamma^{-k}\Gamma^{-l}\rangle
=1.
\]
Normal ordering
\begin{eqnarray}
d^{\alpha}d_{\alpha}(z)&=& p^{\alpha}p_{\alpha}(z)
+p^{\alpha}\sigma^m_{\alpha\dot{\alpha}}\bar{\theta}^{\dot{\alpha}}i\partial
X_m(z) +\frac{1}{8}\partial (\bar{\theta})^2
\bar{\theta}_{\dot\alpha}\bar{\sigma}^{m\dot\alpha\alpha}
\theta_{\alpha}i\partial X_m(z) \nonumber \\
&&+\frac{1}{64}(\theta)^2\partial(\bar{\theta})^2\partial(\bar{\theta})^2(z)
-\frac{1}{2}(\bar{\theta})^2:p^{\alpha}\partial
\theta_{\alpha}(z): +\frac{1}{4}\partial
(\bar{\theta})^2:p^{\alpha}\theta_{\alpha}(z):
\nonumber \\
&&+\frac{1}{4}(\bar{\theta})^2:\partial X^m\partial X_m(z):
-\frac{1}{2}\partial \bar{\theta}_{\dot\alpha}
\partial
\bar{\theta}^{\dot\alpha}(z) +\frac{1}{2}\partial^2
\bar{\theta}_{\dot\alpha}\bar{\theta}^{\dot\alpha}(z).
\end{eqnarray}
The Hermiticity property of the hybrid variables\\
\begin{equation}
\begin{array}{rcl}
(X^m)^{\dagger} &=& X^m,\\
\theta^{\dagger}_{\alpha} &=&
\bar{\theta}_{\dot\alpha}, \\
p_{\alpha}^{\dagger}     &=&
-\bar{p}_{\dot\alpha} ,\\
(\Gamma^{-j})^{\dagger} &=& \frac{1}{2}
\epsilon^{jkl}\Gamma^{-k}\Gamma^{-l}e^{-\rho},
\\
(\Gamma^{+j})^{\dagger}    &=& \frac{1}{2}
\epsilon^{jkl}\Gamma^{+k}\Gamma^{+l}e^{\rho}
,\\
(e^{\rho})^{\dagger} &=& \frac{1}{6}
e^{-2\rho}\epsilon^{jkl}\Gamma^{-j}\Gamma^{-k}\Gamma^{-l},
\\
(X^{-j})^{\dagger} &=& X^{-j} ,\\
(X^{+j})^{\dagger} &=& X^{+j}.
\end{array}
\end{equation}
\end{appendix}

\newpage

\end{document}